\documentclass[aps,prl,twocolumn,groupedaddress,showpacs,onlinecite]{revtex4-1}

\usepackage{color}
\usepackage{graphicx}
\usepackage{dcolumn}
\usepackage{bm}
\usepackage{float}
\usepackage[mathlines]{lineno}
\usepackage{tabularx}
\usepackage{footnote}
\usepackage{amsmath}
\usepackage{txfonts}

\begin{document}

\title{Strong coupling superconductivity in trilayer film LiB$_2$C$_2$}

\author{Miao Gao$^{1,2}$}\email{gaomiao@nbu.edu.cn}
\author{Xun-Wang Yan$^{3}$}
\author{Zhong-Yi Lu$^{4}$}
\author{Tao Xiang$^{5,6}$}

\date{\today}

\affiliation{$^{1}$Department of Microelectronics Science and Engineering, School of Physical Science and Technology, Ningbo University, Zhejiang 315211, China}

\affiliation{$^{2}$Laboratory of Clean Energy Storage and Conversion, Ningbo University, Zhejiang 315211, China}

\affiliation{$^{3}$College of Physics and Engineering, Qufu Normal University, Shandong 273165, China}

\affiliation{$^{4}$Department of Physics, Renmin University of China, Beijing 100872, China}

\affiliation{$^{5}$Institute of Physics, Chinese Academy of Sciences, Beijing 100190, China }

\affiliation{$^{6}$School of Physics, University of Chinese Academy of Sciences}

\begin{abstract}

Coupling between $\sigma$-bonding electrons and phonons is generally very strong. To metallize $\sigma$-electrons provides a promising route to hunt for new high-T$_c$ superconductors.
Based on this picture and first-principles density functional calculation with Wannier interpolation for electronic structure and lattice dynamics, we predict that trilayer film LiB$_2$C$_2$ is a good candidate to realize this kind of high-T$_c$ superconductivity.
By solving the anisotropic Eliashberg equations, we find that free-standing trilayer LiB$_2$C$_2$ is a phonon-mediated superconductor with T$_c$ exceeding the liquid-nitrogen temperature at ambient pressure.
The transition temperature can be further raised to 125 K by applying a biaxial tensile strain.
\end{abstract}

\maketitle

Boosting superconducting transition temperature, T$_c$, is one of the most important goals in the study of high-T$_c$ superconductivity.
According to Bardeen-Cooper-Schrieffer (BCS) theory \cite{Bardeen-PR108}, large density of states (DOS) at the Fermi level, strong electron-phonon coupling (EPC), and high-frequency phonons are beneficial for superconductivity.
These three conditions are simultaneously fulfilled in MgB$_2$, whereas the coupling between metallic covalent $\sigma$ bands and bond-stretching boron phonons play an essential role in its 39 K superconductivity \cite{MgB2,An-PRL86,Y.Kong-PRB64,Yildirim-PRL87,Choi-PRB66,Choi-Nature418}.
Recently, Q-carbon by substituting borons for 27\% carbons was successfully synthesized \cite{Bhaumik-ACSNano11}. This compound shows 55 K superconductivity at ambient pressure, breaking the record of T$_c$ set by MgB$_2$, for purely phonon-mediated superconductors \cite{Bhaumik-ACSNano11}.

To search for new phonon-mediated superconductors with higher T$_c$ at ambient pressure, a number of candidates have been suggested.
Among them, quasi-two-dimensional compounds composed of Li, B, and C, have been studied most intensively.
The parent compound of these materials, LiBC, is a semiconductor, which is isostructural and isovalent to MgB$_2$ \cite{Worle-ZAAC621,Karimov-JPCM16}.
By introducing vacancies at Li sites, Rosner \emph{et al.} suggested that the covalent $\sigma$ bands of LiBC will be partially occupied and become superconducting at about 100 K \cite{Rosner-PRL88}.
Similar prediction was made by Dewhurst \emph{et al.} for Li$_{0.125}$BC \cite{Dewhurst-PRB68}.
However, no evidence of superconductivity was reported down to 2 K in Li-deficient LiBC, i.e. Li$_x$BC,  \cite{Bharathi-SSC124,Souptela-SSC125,Fogg-PRB67,Fogg-CC12}, due to dramatic structural distortions to the boron-carbon layers introduced by Li vacancies, which impedes the metallization of $\sigma$-bonding electrons \cite{Fogg-JACS128}.

Thus a unabridged Li lattice is importance in stablizing the crystal structure. In order to dope holes without introducing lattice distortion, replacing partially carbons by borons was proposed \cite{Miao-JAP113,Gao-PRB91}.
In particular, based on first-principles calculations, we predicted that both Li$_3$B$_4$C$_2$ and Li$_2$B$_3$C could become superconducting above 50 K \cite{Gao-PRB91}.
A similar compound Li$_4$B$_5$C$_3$ was also predicted to be a superconductor at 16.8 K \cite{Bazhirov-PRB89}.
However, to synthesize these B-enriched stoichiometric compounds is difficult \cite{Milashius-ICF5}.
Experimentally, it was reported that hole-doped Li$_x$B$_{1.15}$C$_{0.85}$ shows a drastic decrease in
resistivity below 20 K, but remains non-superconducting \cite{Noguchi-JPCS150}.

\begin{figure}[b]
\begin{center}
\includegraphics[width=8.6cm]{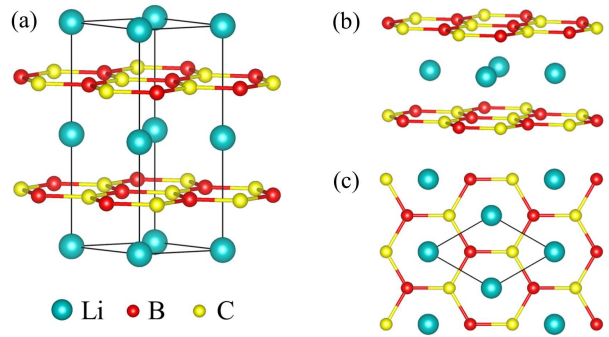}
\caption{Crystal structure of bulk LiBC (a) and that of trilayer LiB$_2$C$_2$ (b). (c) Top view of trilayer LiB$_2$C$_2$. The black line denotes the unit cell. }
\label{fig:Structure}
\end{center}
\end{figure}

Besides doping holes, applying pressure is another way to metallize LiBC.
It was found that the crystal structure of LiBC remains unchanged up to 60 GPa \cite{Lazicki-PRB75}.
Theoretically, the metallization occurs at a calculated pressure of 345 GPa, but the covalent $\sigma$ bands remain unconducting \cite{Lazicki-PRB75}.
By utilizing particle swarm optimization technique, Zhang uncovered a first-order phase transition for LiBC from the low-pressure to a high-pressure insulating phase, at about 108 GPa \cite{Zhang-EPL114}. This transition is accompanied by the formation of $sp^3$-like boron-carbon networks.
Thus, high pressure still can not metallize the covalent $\sigma$ bands of LiBC effectively.

\begin{figure}[t]
\begin{center}
\includegraphics[width=8.6cm]{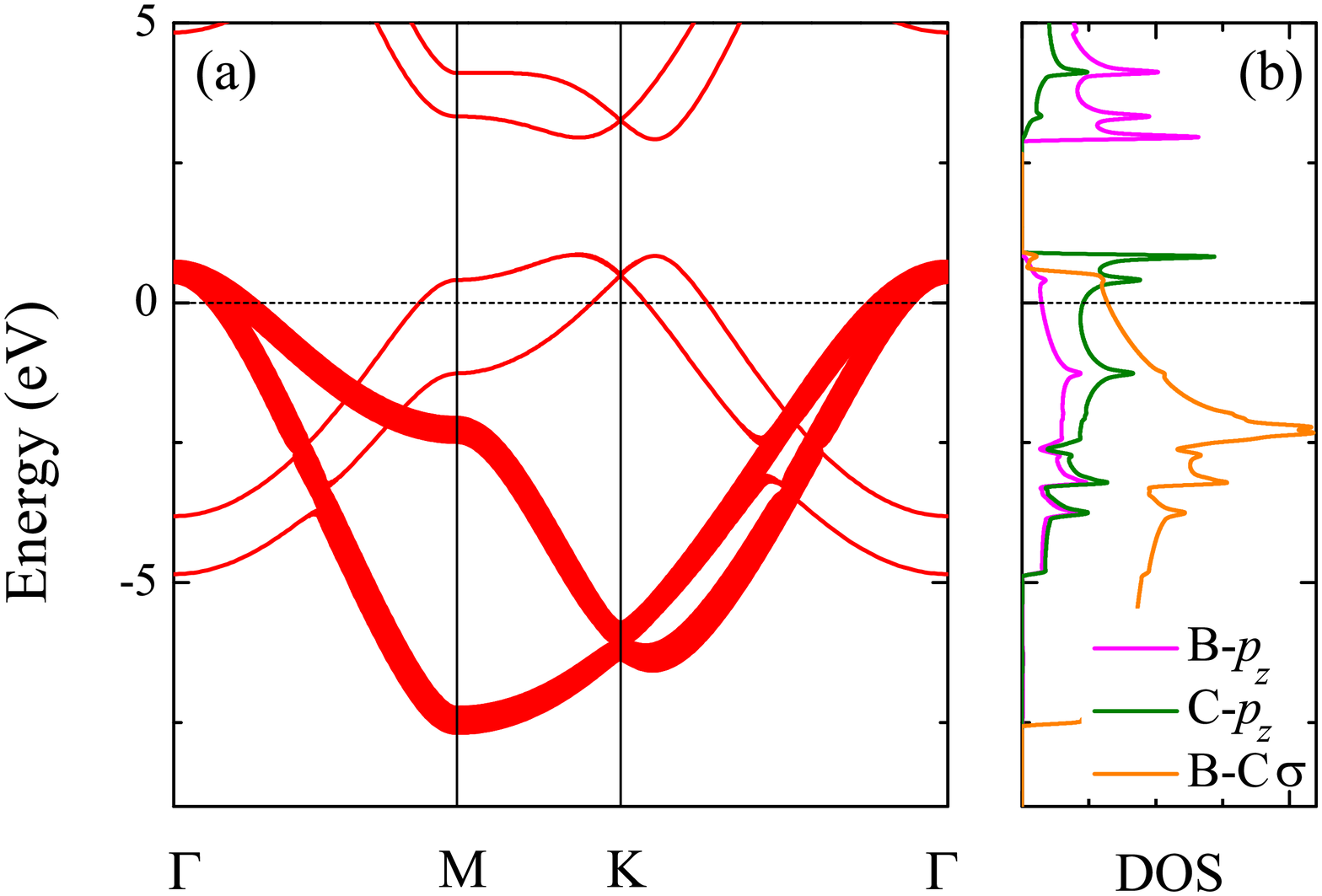}
\includegraphics[width=8.6cm]{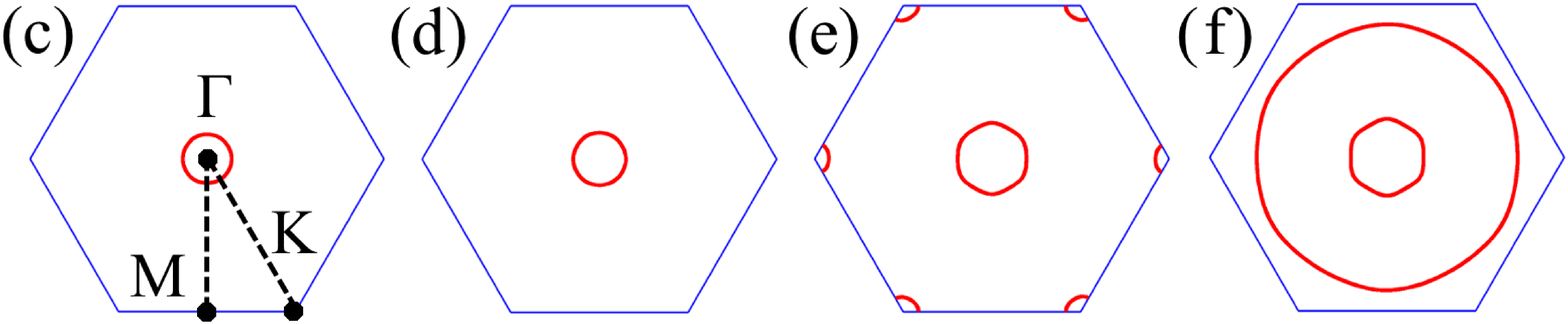}
\caption{Electronic structure of trilayer LiB$_2$C$_2$. (a) Band structure. The width of red line is proportional to weight of $sp^2$-hybridized $\sigma$ orbitals in that band. The Fermi energy was set to zero. (b) Orbital-resolved DOS. (c)-(f) Fermi surfaces.}
\label{fig:Band}
\end{center}
\end{figure}

Is it possible to find a metallic LiBC-like compound which is relatively simple to synthesize?
In this work, we point out that a trilayer LiB$_2$C$_2$ film, which contains two honeycomb boron-carbon sheets intercalated by a vacancy-free triangular Li layer [Fig.~\ref{fig:Structure}], is just such a candidate. As no vacancies or substitutions are involved in trilayer LiB$_2$C$_2$, the holonomic Li lattice can inhibit the structural distortion in boron-carbon sheet. Furthermore, trilayer LiB$_2$C$_2$ is directly derived from bulk LiBC,
it has a high probability to be successfully grown.

We have carried out first-principles calculation in conjunction with the Wannier interpolation technique to determine the electronic structure, lattice dynamics, and EPC for trilayer LiB$_2$C$_2$ \cite{Supp}. We find that both $\sigma$ and $\pi$ bands emerge at the Fermi level in this two-dimensional material. The bond-stretching $E_u$ and $E_g$ phonon modes couple strongly with the metallized $\sigma$ electrons.
After solving the anisotropic Eliashberg equations, we find that the free-standing trilayer LiB$_2$C$_2$ is a two-gap superconductor, with T$_c$ about 92 K.
The superconducting temperature is enhanced by applying a biaxial tensile strain (BTS) to the film.
The optimal BTS appears around 6\%-8\%, at which the transition temperature could even reach 125 K.
This enhancement can be understood by the increase of density of states (DOS) at the Fermi level and the strong softening relevant phonon modes under BTS.

In LiBC, Li atoms occupy the interstitial sites of layered honeycomb boron-carbon sheets [Fig.~\ref{fig:Structure}(a)]. The optimized lattice constants of bulk LiBC, obtained from our calculations, are 2.743 {\AA} and 7.029 {\AA} along the $a$ and $c$ axes, in good agreement with the experimental results ($a$=2.752 {\AA} and $c$=7.058 {\AA}) \cite{Worle-ZAAC621}.
Trilayer LiB$_2$C$_2$ is built from the middle three layers of bulk LiBC [Fig.~\ref{fig:Structure}(b)]. The $c$-axis lattice parameter of the slab model for this trilayer LiB$_2$C$_2$ was set to 15 {\AA} to avoid
unphysical interactions between $c$-axis replicas.
The in-plane lattice constant of trilayer LiB$_2$C$_2$ is 2.706 {\AA}, slightly smaller than in the bulk.
Boron atom moves outward by 0.054 {\AA} with respect to the location of carbon atom, forming a buckled layer with carbons.

\begin{figure}[t]
\begin{center}
\includegraphics[width=8.6cm]{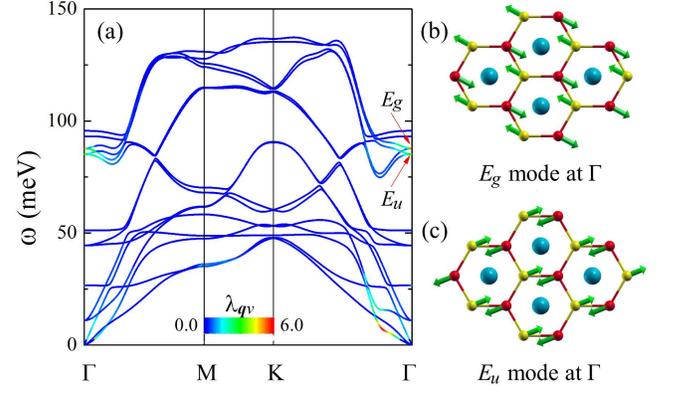}
\caption{(a) Phonon spectrum of trilayer LiB$_2$C$_2$ with a color representation of $\lambda_{{\mathbf q}\nu}$ at given wave vectors and modes.
The vibrational patterns for strongly coupled (b) $E_u$ and (c) $E_g$ phonon modes.
The direction and relative amplitude of atomic movement were represented by green arrows and their lengthes, respectively.}
\label{fig:phonon}
\end{center}
\end{figure}

Figure \ref{fig:Band} shows the band structure, DOS, and Fermi surfaces of trilayer LiB$_2$C$_2$. The $\sigma$ bands are partially filled [Fig.~\ref{fig:Band}(a)]. The $\pi$ (bonding) and $\pi^*$ (antibonding) bands formed by $p_z$ orbitals of carbon and boron atoms are
separated by a direct energy gap of 2.1 eV.
Two Dirac-cone states at the $K$ point are observed.
In comparison with the B-$p_z$ orbital, the C-$p_z$ orbital has larger contribution to the $\pi$ bands [Fig.~\ref{fig:Band}(b)], due to its lower on-site energy.
There are four bands across the Fermi level. The Fermi surface sheets, represented by the two circles [Fig.~\ref{fig:Band}(c) and Fig.~\ref{fig:Band}(d)] and two hexagons [Fig.~\ref{fig:Band}(e) and Fig.~\ref{fig:Band}(f)] surround the $\Gamma$ point, are contributed mainly by $\sigma$ electrons.
The pockets at the Brillouin-zone corners [Fig.~\ref{fig:Band}(e)] and the bigger circle [Fig.~\ref{fig:Band}(f)]
are associated with the $\pi$ bands. The DOS at the Fermi level of trilayer LiB$_2$C$_2$ [Table I] is almost twice that of MgB$_2$ \cite{Y.Kong-PRB64}.

Figure \ref{fig:phonon} shows the $\lambda_{{\bf q}\nu}$-weighted phonon spectrum and vibrational patterns of strongly coupled phonon modes in trilayer LiB$_2$C$_2$.
The free-standing trilayer LiB$_2$C$_2$ is dynamically stable because there is no imaginary frequency in the phonon spectrum [Fig.~\ref{fig:phonon}(a)].
The two strongly coupled phonon modes, $E_u$ and $E_g$, only involve the in-plane vibrations of boron and carbon atoms [Fig.~\ref{fig:phonon}(b) and Fig.~\ref{fig:phonon}(c)].
The frequencies of $E_u$ and $E_g$ modes are respectivly 85.16 meV and 87.68 meV at the $\Gamma$ point, about 20.3\% and 23.8\% higher than that of $E_{2g}$ modes in MgB$_2$ \cite{Bohnen-PRL86}.
This can be attributed to the stronger boron-carbon $\sigma$ bonds and larger interatomic force constants in trilayer LiB$_2$C$_2$.
Besides these two modes, several low-frequency phonon modes have also sizeable contribution to $\lambda_{{\bf q}\nu}$.

\begin{figure}[t]
\begin{center}
\includegraphics[width=8.6cm]{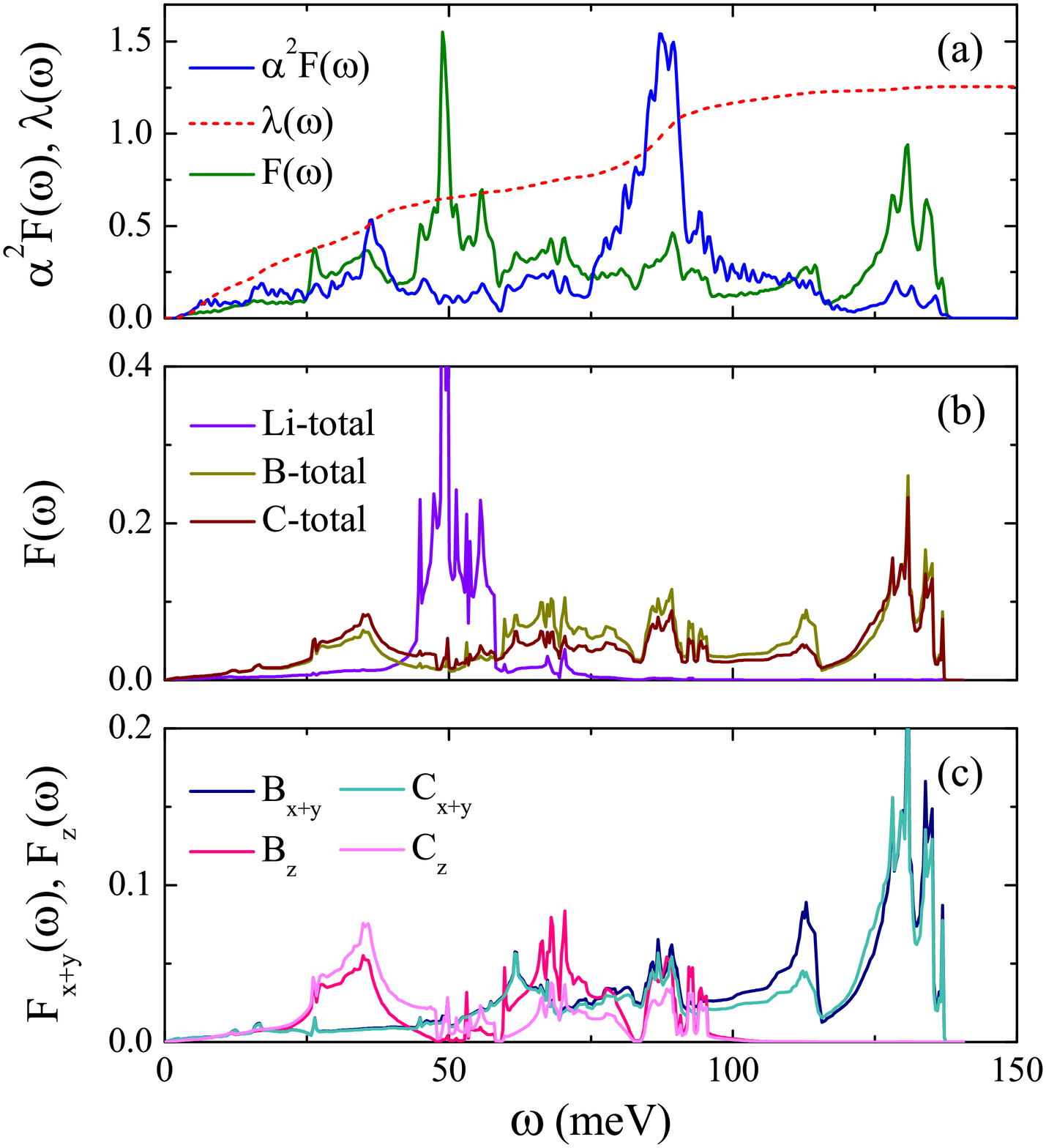}
\caption{(a) Eliashberg spectral function $\alpha^2F(\omega)$, phonon DOS $F(\omega)$, and accumulated $\lambda(\omega)$ for trilayer LiB$_2$C$_2$. (b) Projected phonon DOS.
(c) In-plane and out-of-plane decompositions of boron and carbon phonon DOS.}
\label{fig:a2f}
\end{center}
\end{figure}

Figure \ref{fig:a2f} shows the isotropic Eliashberg spectral function $\alpha^2F(\omega)$, total and projected phonon DOS.
The main peak of $\alpha^2F(\omega)$ around 87 meV results from the $E_u$ and $E_g$ modes.
From the projected phonon DOS calculated through quasi-harmonic approximation [Fig.~\ref{fig:phonon}(b) and Fig.~\ref{fig:phonon}(c)], it is clear that the low-frequency phonon DOS is contributed mainly by
the out-of-plane displacements of boron and carbon atoms.
These modes become more active in EPC due to the removal of quantum confinement \cite{Profeta-NatPhys8,Gao-arXiv}.
A sharp peak of $F(\omega)$ surrounding 50 meV is contributed by Li phonons. However, $\alpha^2F(\omega)$ is insignificant near 50 meV, indicating that the coupling between
electrons and Li phonons is rather weak. The EPC constant $\lambda$ of free-standing trilayer LiB$_2$C$_2$ is 1.25, about 67.1\% higher than that of MgB$_2$ \cite{Margine-PRB87,Eiguren-PRB78,Calandra-PRB82}.

\begin{figure}[t]
\begin{center}
\includegraphics[width=8.6cm]{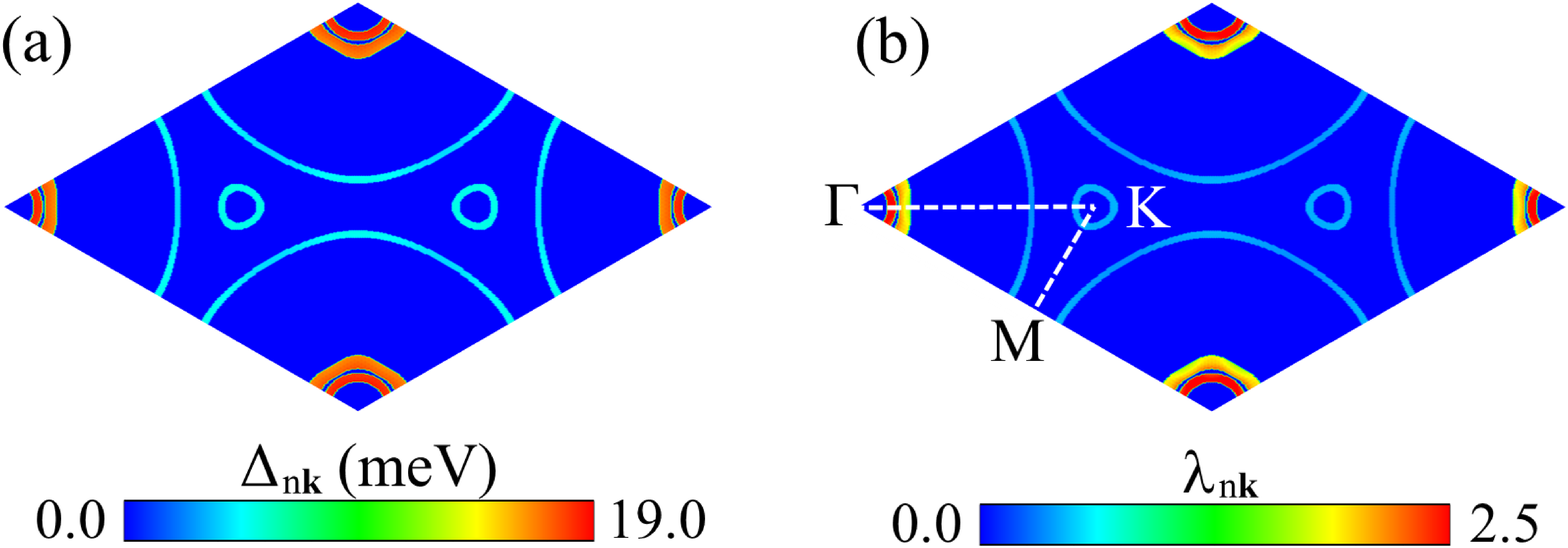}
\includegraphics[width=8.6cm]{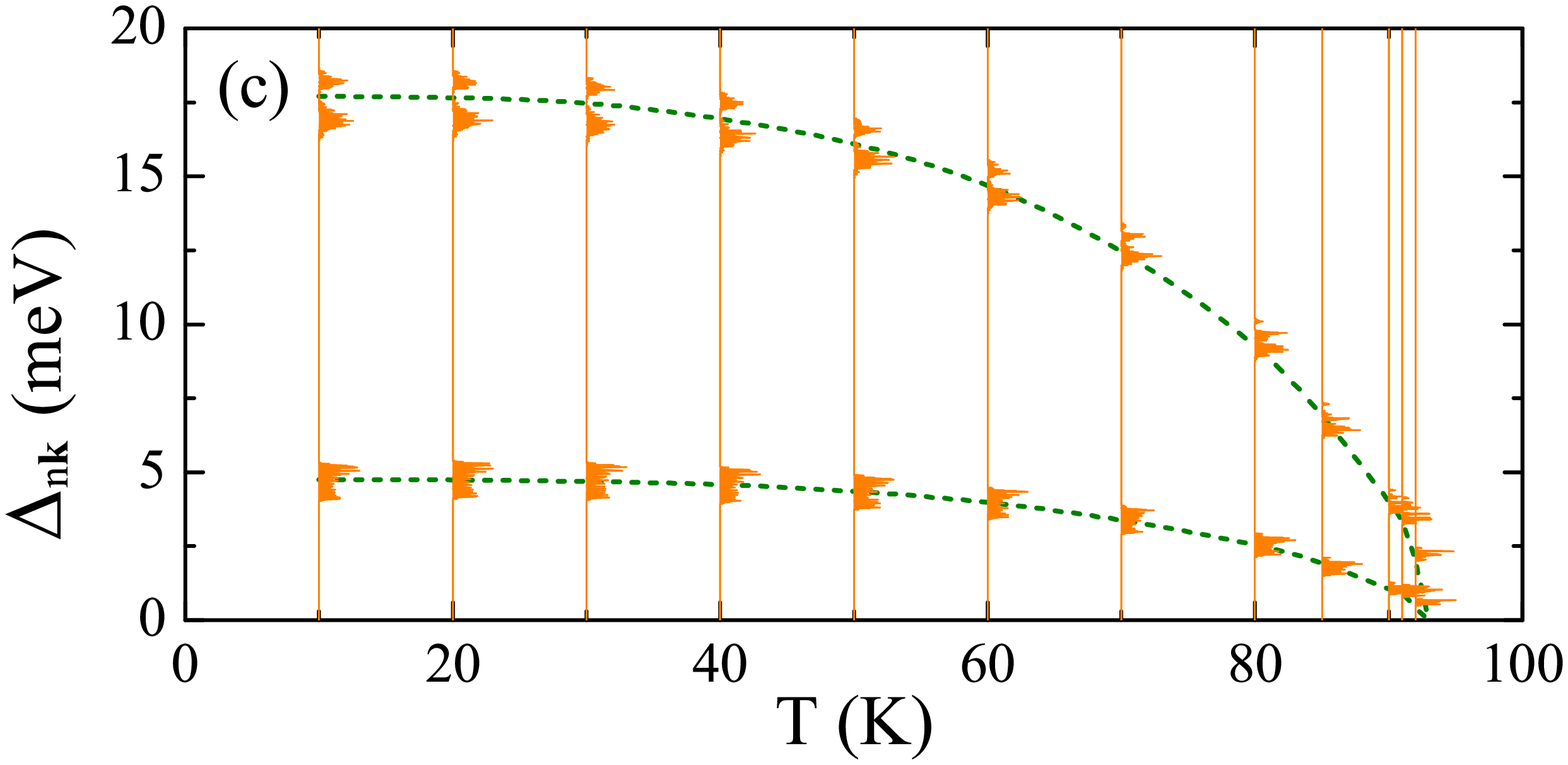}
\caption{(a) Distribution of superconducting gap $\Delta_{n{\bf k}}$ on the Fermi surface at 10 K.
(b) The momentum-resolved EPC strength $\lambda_{n{\bf k}}$ for each electronic state {n\bf k} on the Fermi surface.
Here $\lambda_{n{\bf k}}$ is computed through $\sum_{m{\bf k'}}\lambda(n{\bf k},m{\bf k'},0)\delta({\epsilon_{m{\bf k'}}})/N(0)$ \cite{Margine-PRB87}.
For convenience, these two figures were drawn in the reciprocal unit cell.
Electronic state {n\bf k}, whose energy lies within $\pm$0.1 eV from the Fermi level, are included in the calculation.
(c) Temperature dependence of the gap values $\Delta_{n{\bf k}}$ on the Fermi surface at different temperatures.}
\label{fig:gap}
\end{center}
\end{figure}

\begin{table*}
\begin{center}
\caption{Calculated lattice parameter, electronic structure, EPC properties, and T$_c$ for trilayer LiB$_2$C$_2$ under BTS.
$h_\text{C}$ represents the height of carbon atom from the Li layer. $h_\text{B}$-$h_\text{C}$ is the buckling height of the boron-carbon sheet.
$N_\sigma$(0) and $N_\pi$(0) denote the $\sigma$-band and $\pi$-band DOS at the Fermi level, respectively.
$E_{\sigma_1\Gamma}$ and $E_{\sigma_2\Gamma}$ stand for the energies of valance bands at the $\Gamma$ point.
The frequencies of strongly coupled phonon modes are labelled by $\omega_{E_u}$ and $\omega_{E_g}$.
$\omega _{\text{log}}$ and $\langle\omega^2\rangle$ can be determined through $\exp \left[ \frac{2}{\lambda }\int \frac{d\omega }{%
\omega }\alpha ^{2}F(\omega )\ln \omega \right]$ and $\frac{2}{\lambda}\int d\omega\alpha^2F(\omega)\omega$.
$T_{c,0.1}^{\text{Aniso}}$ and $T_{c,0.2}^{\text{Aniso}}$
are T$_c$s determined by solving the anisotropic Eliashberg equations, when setting the Coulomb pseudopotential $\mu_c^*$ to 0.1 and 0.2.
$T_{c, 0.1}^{\text{MAD}}$ stands for T$_c$ evaluated by the semiempirical McMillian-Allen-Dynes formula \cite{Allen-RPB12_905} with $\mu_c^*$ of 0.1.
The units for height, DOS, energy, frequency, and T$_c$ are {\AA}, states$\cdot$spin$^{-1}$$\cdot$eV$^{-1}$$\cdot$cell$^{-1}$, eV, meV, and K, respectively.}
\label{table:energy}
\begin{tabular}{p{0.5cm}p{1.0cm}p{1.0cm}p{1.0cm}p{1.0cm}p{1.0cm}p{1.0cm}p{1.0cm}p{1.0cm}p{1.0cm}p{1.0cm}p{1.0cm}p{1.0cm}p{1.0cm}p{1.0cm}p{1.0cm}}
  \hline
  \hline
  $\epsilon$ & $h_\text{C}$ & $h_\text{B}$-$h_\text{C}$ & $N$(0) & $N_\sigma$(0) & $N_\pi$(0) & $E_{\sigma_1\Gamma}$ & $E_{\sigma_2\Gamma}$ & $\omega_{E_u}$ & $\omega_{E_g}$ &$\lambda$ & $\omega_{\text{log}}$ & $\sqrt{\langle\omega^2\rangle}$ & $T_{c, 0.1}^{\text{MAD}}$ & $T_{c,0.1}^{\text{Aniso}}$ & $T_{c,0.2}^{\text{Aniso}}$ \\
  \hline
  0  & 1.721 & 0.054 & 0.602 & 0.319 & 0.282 & 0.527  & 0.578  & 85.16 & 87.68 & 1.25 & 38.60 & 63.37 & 42.3 &  92 & 82 \\
  2  & 1.710 & 0.050 & 0.613 & 0.323 & 0.290 & 0.420  & 0.479  & 69.14 & 72.43 & 1.19 & 46.57 & 63.26 & 47.8 & 102 & 95 \\
  4  & 1.696 & 0.048 & 0.624 & 0.326 & 0.298 & 0.317  & 0.386  & 52.25 & 56.62 & 1.30 & 49.15 & 58.91 & 55.9 & 115 & 108 \\
  6  & 1.680 & 0.045 & 0.634 & 0.328 & 0.306 & 0.218  & 0.298  & 36.22 & 42.52 & 1.60 & 44.59 & 51.68 & 70.9 & 125 & 119 \\
  8  & 1.661 & 0.043 & 0.643 & 0.330 & 0.312 & 0.121  & 0.214  & 32.87 & 39.80 & 1.52 & 47.62 & 51.99 & 71.0 & 125 & 120 \\
 10  & 1.639 & 0.041 & 0.624 & 0.306 & 0.318 & 0.022  & 0.133  & 47.95 & 51.80 & 0.95 & 52.42 & 59.24 & 39.3 & 102 &  97 \\
 12  & 1.618 & 0.040 & 0.493 & 0.171 & 0.323 & -0.097 & 0.033  & 70.01 & 71.10 & 0.41 & 47.15 & 63.39 & 2.5 & -- & --\\
 14  & 1.604 & 0.038 & 0.336 & 0.010 & 0.326 & -0.247 & -0.101 & 84.21 & 84.06 & 0.30 & 43.70 & 53.35 & 0.2 & -- & --$^\text{a}$\\
  \hline
  \hline
\end{tabular}
\end{center}
\begin{flushleft}
\footnotesize{$^\text{a}$ When the EPC is weak, it is very difficult to ascertain the exact value of T$_c$ for trilayer LiB$_2$C$_2$ at low temperature
due to prohibitive computational cost.}\\
\end{flushleft}
\end{table*}

Figure \ref{fig:gap} shows the distribution of superconducting energy gaps $\Delta_{n{\bf k}}$ and  $\lambda_{n{\bf k}}$ on the Fermi surface.
There are two anisotropic superconducting gaps, associated with the $\sigma$ bands and the $\pi$ bands [Fig.~\ref{fig:gap}(a)], respectively.
The highest temperature with non-vanished gap is about 92 K, corresponding to T$_c$ [Fig.~\ref{fig:gap}(c)].
The two-gap superconductivity results from the anisotropy of EPC constant $\lambda_{n{\bf k}}$ on different Fermi sheets [Fig.~\ref{fig:gap}(b)].
The $\sigma$ electrons, especially those around the inner pocket shown in [Fig.~\ref{fig:gap}(b)], couple strongly with the $E_u$ and $E_g$ modes.
The Fermi-surface averaged gaps are $\Delta_\sigma$=17.3 meV and $\Delta_\pi$=4.7 meV at 10 K. The anisotropy of $\Delta_\sigma$
is slightly stronger than that of $\Delta_\pi$ [Fig.~\ref{fig:gap}(c)].
In MgB$_2$, the measured $\Delta_\sigma$ and $\Delta_\pi$ at 4.2 K are in the ranges of 7.0-7.1 meV and 2.3-2.8 meV \cite{Iavarone-PRL89,Szabo-PRL87,Gonnelli-PRL89}, respectively.
$\Delta_\sigma$ of trilayer LiB$_2$C$_2$ is about 2.47 times that of MgB$_2$.

To further raise T$_c$, we apply a BTS, described by $\epsilon =(a-a_0)/a_0\times100\%$, to trilayer LiB$_2$C$_2$.
Here $a_0$ and $a$ are the in-plane lattice constants before and after BTS.
The boron-carbon sheet becomes more and more flat with the increase of BTS,
and the separation between two layers is gradually depressed [Table I].
These structural changes have significant impacts on the band structure of trilayer LiB$_2$C$_2$.
On one hand, the $p_z$ orbital experiences an enhanced Coulomb repulsion from Li layer.
As a consequence, the energies of $\pi$ bands are increased.
The valence bands near the $\Gamma$ point shrink below the Fermi level, but the energy of Dirac point is almost unaffected with respect to the Fermi level.
On the other hand, BTS reduces the overlap among atomic orbitals and weakens the dispersion of energy bands, enlarging the electronic DOS at the Fermi level.
Above 12\% BTS, there is a sudden abatement in the $\sigma$-band DOS at the Fermi level, $N_\sigma$(0), due to almost complete occupation of two $\sigma$ bands  [Table I].

\begin{figure}[b]
\begin{center}
\includegraphics[width=8.6cm]{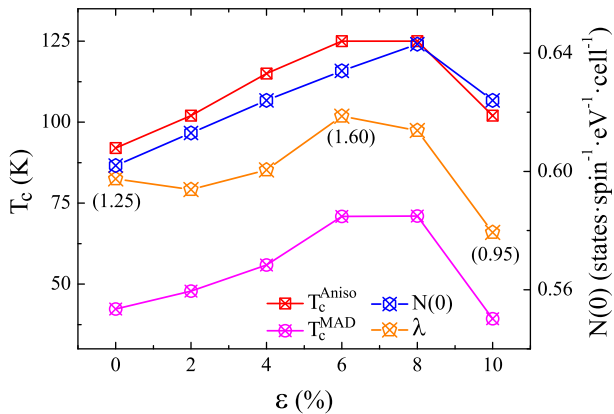}
\caption{The T$_c$, $N(0)$, and $\lambda$ of trilayer LiB$_2$C$_2$ under BTS. The T$_c$ obtained by the McMillian-Allen-Dynes formula is also given for comparison.}
\label{fig:Tc}
\end{center}
\end{figure}

With the increase of BTS, the strongly coupled phonon modes become softened [Table I], and the EPC constant $\lambda$ is increased.
Compared with the strain-free case, $\lambda$ increases by 28.0\% under 6\% BTS.
Moreover, trilayer LiB$_2$C$_2$ is rather stable against BTS. No imaginary phonon frequency is found up to 8\% BTS.
A tiny imaginary phonon frequency of 1.97$i$ meV emerges only when BTS is above 14\%.
Even this imaginary frequency may not be a signature of lattice instability, because similar phenomenon was also found in the simulations of germanene \cite{Cahangirov-PRL102}, $\beta_{12}$ borophene \cite{Gao-PRB95}, buckled arsenene \cite{Kamal-PRB91,Kong-CPB27}.
This phenomenon may result from numerical difficulties in determining rapid decayed interatomic forces \cite{Sahin-PRB80}.
Above 10\% BTS, there are abnormal arises of phonon frequencies for the $E_u$ and $E_g$ modes [Table I], probably related to the decline of $N_\sigma(0)$.

Figure \ref{fig:Tc} shows the BTS dependence of T$_c$, determined by by self-consistently solving the anisotropic Eliashberg equations.
A dome-like structure is observed. The maximal T$_c$ is about 125 K.
For strained trilayer LiB$_2$C$_2$, the transition temperature is predominantly determined by the electronic DOS at the Fermi level and the EPC constant $\lambda$ [Fig.~\ref{fig:Tc}].

Since mechanical exfoliation from the bulk phase is a robust method to produce ultraclean, highly crystalline thin films \cite{Geim-NatMater6},
we also examine the possibility of synthesizing trilayer LiB$_2$C$_2$ from LiBC.
After cleaving the (0001) plane of LiBC, we find that the most favorable structure is a half-Li-terminated surface, with evenly distributed Li atoms.
This can balance the chemical valence as uniformly as possible.
A BC-sheet-terminated surface has a disadvantage in energy, about 0.092 eV/${\AA}^2$.
So the film that we can obtain after exfoliation is not trilayer LiB$_2$C$_2$, but trilayer LiB$_2$C$_2$ with half-Li covering on each side.
The exfoliation energy for half-Li-covered trilayer LiB$_2$C$_2$ is 0.142 eV/${\AA}^2$, about six times that of graphene \cite{Supp}.
The extra surface Li on trilayer LiB$_2$C$_2$ can be further removed, for example, through vertical manipulation of tip in scanning tunneling
microscopy (STM) experiment \cite{Hla-STM}.

In summary, based on first-principles density functional calculation and Wannier interpolation, we show that the $\sigma$-bands in trilayer LiB$_2$C$_2$ are partially occupied.
These metallized $\sigma$ electrons couple strongly with the $E_u$ and $E_g$ phonon modes, driving this material into a high-T$_c$ superconducting phase at ambient pressure.
Applying biaxial tensile strain to trilayer LiB$_2$C$_2$ can significantly boost the T$_c$ to a higher temperature.

M. G. thank E. R. Margine for her help in the calculation and W. Ji for useful discussion.
This work was supported by the National Key R \& D Program of China
(Grant No. 2017YFA0302900),
National Natural Science Foundation of China (Grant Nos. 11888101, 11774422, 11974194, 11974207),
and Zhejiang Provincial Natural Science Foundation of China (Grant No. LY17A040005).
M.G. was also sponsored by K. C. Wong Magna Fund in Ningbo University.


\begin{references}

\bibitem{Bardeen-PR108}J. Bardeen, L. N. Cooper, and J. R. Schrieffer,
Phys. Rev. {\bf 108}, 1175 (1957)

\bibitem{MgB2} J. Nagamatsu, N. Nakagawa, T. Muranaka, Y. Zenitani, and J. Akimitsu,
Nature (London) {\bf 410}, 63 (2001).

\bibitem{An-PRL86}J. M. An and W. E. Pickett,
Phys. Rev. Lett. {\bf 86}, 4366 (2001).

\bibitem{Y.Kong-PRB64}Y. Kong, O. V. Dolgov, O. Jepsen, and O. K. Andersen,
Phys. Rev. B {\bf 64}, 020501(R) (2001).

\bibitem{Yildirim-PRL87}T. Yildirim \textit{et al.}, Phys. Rev. Lett. {\bf 87}, 037001 (2001).

\bibitem{Choi-PRB66}H. J. Choi, D. Roundy, H. Sun, M. L. Cohen, and S. G. Louie,
Phys. Rev. B {\bf 66}, 020513(R) (2002).

\bibitem{Choi-Nature418}H. J. Choi, D. Roundy, H. Sun, M. L. Cohen, and S. G. Louie,
Nature (London) {\bf 418}, 758 (2002).

\bibitem{Bhaumik-ACSNano11}A. Bhaumik, R. Sachan, S. Gupta, and J. Narayan, ACS Nano {\bf 11}, 11915 (2017).

\bibitem{Worle-ZAAC621}M. W\"{o}rle, R. Nesper, G. Mair, M. Schwarz, and H. G. Vonschnering, Z.
Anorg. Allg. Chem. {\bf 621}, 1153 (1995). 

\bibitem{Karimov-JPCM16}P. F. Karimov, N. A. Skorikov, E. Z. Kurmaev, L. D. Finkelstein, S. Leitch, J. MacNaughton, A. Moewes, and T. Mori,
J. Phys.: Condens. Matter {\bf 16}, 5137 (2004).

\bibitem{Rosner-PRL88} H. Rosner, A. Kitaigorodsky, and W. E. Pickett,
Phys. Rev. Lett. {\bf 88}, 127001 (2002).

\bibitem{Dewhurst-PRB68}J. K. Dewhurst, S. Sharma, C. Ambrosch-Draxl, and B. Johansson,
Phys. Rev. B {\bf 68}, 020504(R) (2003).

\bibitem{Bharathi-SSC124}A. Bharathi, S. J. Balaselvi, M. Premila, T. N. Sairam, G. L. N. Reddy, C. S. Sundar, and Y. Hariharan, Solid State Commun. {\bf 124}, 423 (2002).

\bibitem{Souptela-SSC125}D. Souptela, Z. Hossainb, G. Behra, W. L\"{o}sera, and C. Geibel,
Solid State Commun. {\bf 125}, 17 (2003). 

\bibitem{Fogg-PRB67}A. M. Fogg, P. R. Chalker, J. B. Claridge, G. R. Darling, and M. J. Rosseinsky,
Phys. Rev. B {\bf 67}, 245106 (2003). 

\bibitem{Fogg-CC12}A. M. Fogg, J. B. Claridge, G. R. Darling, and M. J. Rosseinsky,
Chem. Commun. {\bf 12}, 1348 (2003). 

\bibitem{Fogg-JACS128}A. M. Fogg, J. Meldrum, G. R. Darling, J. B. Claridge, and
M. J. Rosseinsky, J. Am. Chem. Soc. {\bf 128}, 10043 (2006). 

\bibitem{Miao-JAP113}R. Miao, J. Yang, M. Jiang, Q. Zhang, D. Cai, C. Fan, Z. Bai, C. Liu, F. Wu, and S. Ma, J. Appl. Phys. {\bf 113}, 133910 (2013).

\bibitem{Gao-PRB91}M. Gao, Z.-Y. Lu, and T. Xiang, Phys. Rev. B {\bf 91}, 045132 (2015).

\bibitem{Bazhirov-PRB89}T. Bazhirov, Y. Sakai, S. Saito, and M. L. Cohen, Phys. Rev. B {\bf 89}, 045136 (2014).

\bibitem{Milashius-ICF5}V. Milashius, V. Pavlyuk, G. Dmytriv, and H. Ehrenberg,
Inorg. Chem. Front, {\bf 5}, 853 (2018).

\bibitem{Noguchi-JPCS150}A. Noguchi, S. Emori, Y. Takahashi, K. Takase,
T. Watanabe, K. Sekizawa, and Y. Takano, J. Phys.: Conf. Ser. {\bf 150}, 052188 (2009).

\bibitem{Lazicki-PRB75}A. Lazicki, C.-S. Yoo, H. Cynn, W. J. Evans, W. E. Pickett, J. Olamit, Kai Liu, and Y. Ohishi,
Phys. Rev. B {\bf 75}, 054507 (2007).

\bibitem{Zhang-EPL114}M. Zhang, EPL {\bf 114}, 16001 (2016).

\bibitem{Supp}See Supplemental Material for the technical details of density functional first-principles calculations and Wannier interpolation,
and necessary formulas.

\bibitem{Bohnen-PRL86}K.-P. Bohnen, R. Heid, and B. Renker, Phys. Rev. Lett. {\bf 86}, 5771 (2001).

\bibitem{Profeta-NatPhys8}G. Profeta, M. Calandra, and F. Mauri,
Nat. Phys. {\bf 8}, 131 (2012).

\bibitem{Gao-arXiv}M. Gao. X.-W. Yan, J. Wang, Z.-Y. Lu, and T. Xiang, Phys. Rev. B {\bf 100}, 024503 (2019).

\bibitem{Margine-PRB87}E. R. Margine and F. Giustino, Phys. Rev. B {\bf 87}, 024505 (2013).

\bibitem{Eiguren-PRB78}A. Eiguren and C. Ambrosch-Draxl, Phys. Rev. B {\bf 78}, 045124 (2008).

\bibitem{Calandra-PRB82}M. Calandra, G. Profeta, and F. Mauri, Phys. Rev. B {\bf 82}, 165111 (2010).

\bibitem{Iavarone-PRL89}M. Iavarone, G. Karapetrov, A. E. Koshelev, W. K. Kwok,
G. W. Crabtree, D. G. Hinks, W. N. Kang, E.-M. Choi, H. J.
Kim, H.-J. Kim, and S. I. Lee, Phys. Rev. Lett. {\bf 89}, 187002 (2002).

\bibitem{Szabo-PRL87}P. Szab\'{o}, P. Samuely, J. Ka\u{c}mar\u{c}\'{i}k, T. Klein, J. Marcus, D. Fruchart,
S. Miraglia, C. Marcenat, and A. G. M. Jansen, Phys. Rev. Lett. {\bf 87}, 137005 (2001).

\bibitem{Gonnelli-PRL89}R. S. Gonnelli, D. Daghero, G. A. Ummarino, V. A. Stepanov,
J. Jun, S. M. Kazakov, and J. Karpinski, Phys. Rev. Lett. {\bf 89}, 247004 (2002).

\bibitem{Cahangirov-PRL102}S. Cahangirov, M. Topsakal, E. Akt\"{u}rk, H. \c{S}ahin, and S. Ciraci,
Phys. Rev. Lett. {\bf 102}, 236804 (2009).

\bibitem{Gao-PRB95}M. Gao, Q.-Z. Li, X.-W. Yan, and J. Wang, Phys. Rev. B {\bf 95}, 024505 (2017).

\bibitem{Kamal-PRB91}C. Kamal and M. Ezawa,
Phys. Rev. B {\bf 91}, 085423 (2015).

\bibitem{Kong-CPB27}X. Kong, M. Gao, X.-W. Yan, Z.-Y. Lu, and T. Xiang, Chin. Phys. B {\bf 27}, 046301 (2018).

\bibitem{Sahin-PRB80}H. \c{S}ahin, S. Cahangirov, M. Topsakal, E. Bekaroglu, E. Akturk, R. T. Senger, and S. Ciraci,
Phys. Rev. B {\bf 80}, 155453 (2009).

\bibitem{Allen-RPB12_905}P. B. Allen and R. C. Dynes, Phys. Rev. B {\bf 12}, 905 (1975).

\bibitem{Geim-NatMater6}A. K. Geim and K. S. Novoselov, Nat. Mater. {\bf 6}, 183 (2007).

\bibitem{Hla-STM}S.-W. Hla, J. Vac. Sci. Tech. {\bf 23}, 1351 (2005).

\end{references}
\end{document}